\documentclass[preprint,showpacs,preprintnumbers,amsmath,amssymb]{revtex4}

\usepackage{graphicx}
\usepackage{dcolumn}
\usepackage{bm}

\begin{document}

\title{Calculation of Helium Ground State Energy by Bohr's Theory-Based Methods}
\author{Youhei Tsubono}
 \email{tubonoy-tky@umin.ac.jp}
\affiliation
{The Institute of Science, The University of Tokyo,  6-1 Shirokanedai4-chome, Minato-ku, Tokyo 108-8639, JAPAN}

\date{\today}

\begin{abstract} 
Bohr's model agreed with the hydrogen spectrum results, but did not agree with the spectrum of Helium. Here we show that Bohr's model-based methods can calculate the experimental value (-79.005 eV) of Helium ground state energy correctly. we suppose the orbital planes of the two electrons are perpendicular to each other. By a computational method, we calculate the Coulomb force among the particles, and the number of de Broglie's waves contained in the short segment at short time intervals. Our results demonstrate that two electrons of Helium are actually moving around, not as electron clouds.
\end{abstract}


\maketitle

In 1913, Niels Bohr postulates the Bohr's model which agreed with the observed hydrogen spectrum \cite{rone}. In 1923 Louis de Broglie suggested that electrons might have wave aspect and its wavelength $\lambda$ is equal to $h/mv$, where $h$ is Plank's 
constant  (= $6.62606896 \times 10^{-34}$ Js) and $m$ is the electron  mass (= $9.1093826 \times 10^{-31}$ kg).
In 1927, Davisson and Germer experimentally confirmed de Broglie's hypothesis \cite{rtwo}. Recently the results of the two-slit experiment of an electron showed its wavelike properties \cite{rthree}.

In the Bohr's theory, the circular orbital length is equal to a integer times the wavelength of the electron, so we have, $2\pi r=n \times h/mv$. In several phenomena, the Bohr's model provides good accuracy \cite{rfour,rfive,rsix}. But this model could not explain about the spin of the electron and the two-electron atoms such as the helium exactly. Because of such problems, the Bohr's theory was replaced by the quantum mechanical theory based on the Schroedinger equation in 1920's. The solution of the Schroedinger wave equation showed that the orbital angular momentum of the electron in the ground state of the hydrogen atom is zero.  And the solution also showed the relation between $L$ (the total angular momentum quantum number) and $M$ (the z component of $L$) 
\begin{equation}
M=-L, -L+1, -L+2 ,~~ \dots ~,~ L-2, L-1, L 
\label{eq:eone}
\end{equation}
Eq. (\ref{eq:eone}) and the results of the Stern-Gerlach experiment indicated that an electron has $\pm \frac{1}{2} \hbar$ spin angular momentum. But we can not visualize the electronic motion concretely. The Coulomb potential may be infinitely negative, when the electron is close to the nucleus. And by equating the angular momentum of the spinnig sphere of the electron to $\pm \frac{1}{2} \hbar$, the sphere speed leads to about one hundred times the speed of light \cite{rseven}. 
To solve the above problem, we try going back to the Bohr's model.  

The helium atom has two electrons and a nucleus of charge +2e. The Schroedinger equation for the helium can not be precisely solved. Using the perturbation method, we can get the ground state energy of the helium, which is very close to the experimental value \cite{reight,rnine,rten,releven}. But the calculation of the high order correction is very difficult. 

In this paper, we try to calculate the ground state energy of the helium atom using the new theory based on the Bohr's model, and check if the calculation value is equal to the experimental value -79.005 eV. 

\begin{figure}
\includegraphics[width=5.5cm]{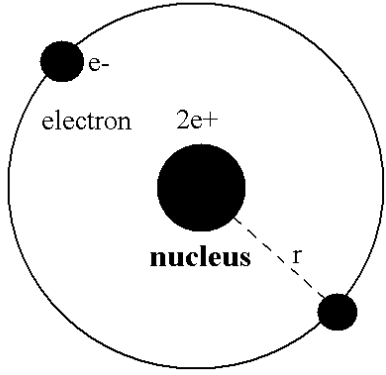}
\caption{\label{fig:fone} One schematic model of Helium in which two electrons are moving on the opposite sides of the nucleus.}
\end{figure}

First, suppose we have one model as shown in Fig.~\ref{fig:fone}. In that model, two electrons of the  helium are on the opposite sides of the nucleus and moving on the same circular orbital.

\vspace{1cm}

Equating the centrifugal force to the Coulomb force, we have
\begin{equation}
\frac{mv^{2}}{r}=\frac{2e^{2}}{4\pi\epsilon r^{2}}-\frac{e^{2}}{4\pi\epsilon (2r)^{2}} 
\label{eq:etwo}
\end{equation}
where $r$ is the circular orbital radius (m), $e$ is the electron charge (= $1.60217653  \times 10^{-19} C$), and $\epsilon$ is the permittivity of vacuum (= $8.854187817 \times  10^{-12} \frac{C^{2}}{Nm^{2}} $)
The circular orbital length is supposed to be an integer times the wavelength of the electron, we have
\begin{equation}
2\pi r=\frac{h}{mv} \times n
\label{eq:ethree}
\end{equation}
The total energy $E$ of the electrons is the sum of the kinetic energy and the Coulomb potential energy, so
\begin{equation}
E=2 \times \frac{mv^{2}}{2}-2 \times \frac{2e^{2}}{4\pi\epsilon r}+\frac{e^{2}}{4\pi\epsilon (2r)}
\label{eq:efour}
\end{equation}
Solving the above three Eqs.~(\ref{eq:etwo}-\ref{eq:efour}), the ground state energy (n=1) is  -83.33 eV. The value is lower than the experimental value -79.005 eV. In this model, as the two electrons are on the same orbital, it is possible that de Broglie's  waves of the two electrons interfere with each other and their motions are affected, so Eqs.~(\ref{eq:etwo}-\ref{eq:efour}) may not be satisfied. Whether the wave is the electron itself or by the field change around the electron can not be ascertained.

\begin{figure}
\includegraphics[width=6cm]{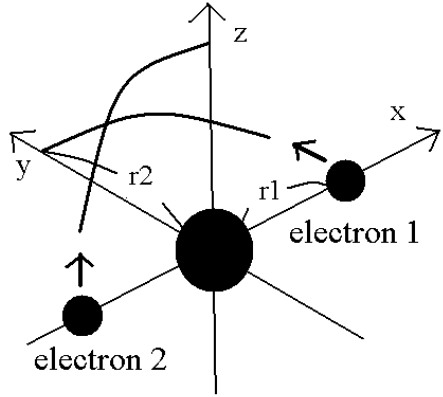}
\caption{\label{fig:ftwo} Schematic model of Helium. Two same-shaped orbital planes are perpendicular to each other. This figure shows one quarter of the orbitals. Electron 1 starts at (r1, 0, 0), while electron 2 starts at (-r1, 0, 0).}
\end{figure}

\begin{figure}
\includegraphics[width=6cm]{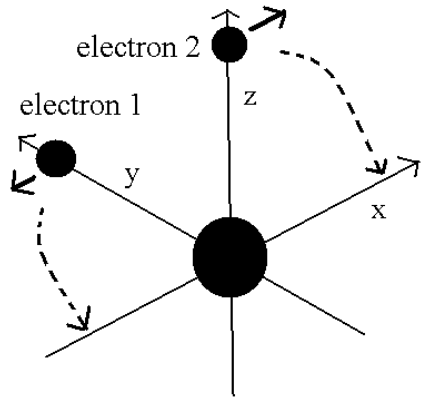}
\caption{\label{fig:fthree} Schematic model of Helium. Two electrons have moved one quarter of their orbitals. Electron 1 is crossing y axis perpendicularly, while electron 2 is crossing z axis.}
\end{figure}

To avoid such problems, we suppose another model as shown in Fig.~\ref{fig:ftwo} and \ref{fig:fthree}. In that model, the planes of the two same-shaped orbitals are perpendicular to each other. As any point on the electron 1 orbital is at the same distance from the points on the both-side ($\pm$ z)  electron 2 orbital, de Bloglie's wave of the electron 2 may interfere with itself and vanish on the electron 1 orbital. So the motion of electron 1 may not  be affected by the wave of electron 2.


Here we investigate how the electrons of the helium are moving by calculating the Coulomb force among the two electrons and the nucleus at short time intervals.
The computer program (class filename: MathMethod) written in the JAVA language (version 1.5.0) to calculate the electronic orbital of the helium is shown in Supplementary Methods.

As shown in Fig.~\ref{fig:ftwo} and  \ref{fig:fthree}, the helium nucleus is at the origin, the orbital plane of the electron 1 is the xy-plane, and that of the electron 2 is xz-plane. The electron 1 initially at (r1, 0, 0) (Fig.~\ref{fig:ftwo}) moves one quarter of its orbital to (0, r2, 0) (Fig.~\ref{fig:fthree}), while the electron 2 initially at (-r1, 0, 0) moves to (0, 0, r2). Meter and second are rather large units for measurement of atomic behavior, here we use new convenient units $MM$ 
 (1 $MM$ = $1 \times 10^{-14}$ meter), $SS$ (1 $SS$ = $1 \times 10^{-22}$ second) and  $MM/SS$ (1 $MM/SS$ = $1 \times 10^{8}$ meter/second).       

In this program, we first input the initial x-coordinate r1 (in $MM$) of the electron 1, and the absolute value of the total energy $E$ (in eV) of the helium. From the inputted value, we calculate the initial velocity of the electron. And at intervals of 1 $SS$ we calculate the Coulomb force among the two electrons and the nucleus.  When the electron 1 is at (xx, yy, 0), the electron 2 is at (-xx, 0, yy) (in $MM$). Change $MM$ to meter as follows; x (m) = xx $\times 10^{-14}$. y (m) = yy $\times 10^{-14}$. So the x component of the acceleration ($m/sec^{2}$) of the electron 1  is
\begin{equation}
-\frac{2e^{2} x}{4\pi\epsilon (rm)(x^{2}+y^{2})^{\frac{3}{2}}}+\frac{e^{2} \times (2x)}{4\pi\epsilon (rm)(4x^{2}+2y^{2})^{\frac{3}{2}}}
\label{eq:efive}
\end{equation}
where the first term is by the Coulomb force between the nucleus and the electron 1, and the second term is by the force between the two electrons. Considering the helium nuclear mass, we use here the reduced mass ($rm$ = 9.1069044 $\times 10^{-31}$ kg) except when the center of mass is at the origin.

In the same way, the y component of the acceleration ($m/sec^{2}$) is 
\begin{equation}
-\frac{2e^{2} y}{4\pi\epsilon (rm)(x^{2}+y^{2})^{\frac{3}{2}}}+\frac{e^{2} \times y}{4\pi\epsilon (rm)(4x^{2}+2y^{2})^{\frac{3}{2}}}
\label{eq:esix}
\end{equation}
Change $m/sec^{2}$ to $MM/SS^{2}$ using the next relation  
\begin{equation}
1 m/sec^{2} = 1 \times 10^{-30} MM/SS^{2}
\label{eq:eseven}
\end{equation}
Based on that calculation value we change the velocity vector and the position of the electrons. We suppose electron 1 moves only on the xy-plane, so the z component of the acceleration of the electron 1 is not considered. We also calculate de Broglie's wavelength of the electron  from the velocity ($\lambda = h/mv$) at intervals of 1 $SS$.  The number of that wave ($\lambda$ in length) contained in that short movement section  (The sum of them is WN) is
\begin{equation}
\frac{\sqrt{VX^{2}+VY^{2}} \times 10^{-14}}{\frac{h}{(rm) \sqrt{VX^{2}+VY^{2}} \times 10^{8} }}
\label{eq:eeight}
\end{equation}
where ($VX, VY$) are the velocity of the electron 1 (in $MM/SS$), the numerator is the movement distance (in meter) for 1 $SS$. the denominator is de Broglie's wavelength (in meter).
Here, the estimated electronic orbital is divided into more than one hundred thousand for the calculation. 
When the electron 1 has moved one quarter of its orbital and its x-coordinate is zero (Fig.~\ref{fig:fthree}), this program checked the y-component of the electron 1 velocity (last $VY$). When the last $VY$ is zero, two electrons are periodically moving around the nucleus on the same orbitals as shown in Fig.~\ref{fig:ftwo} and \ref{fig:fthree}.  So, only when -0.0001 $< VY <$ 0.0001 ($MM/SS$) is satisfied, the program displays the following values  on the screen, r1, $VY$, $preVY$ ($VY$ 1$SS$ ago), and (mid)WN (the total number of de Broglie's waves contained in one quarter of the orbital). The initial inputted x-coordinate of the electron 1 is automatically increased by 1 MM per above calculation to +100 $MM$. 

\begin{table}
\caption{\label{tab:table1}Results of r1 and WN (Number of de Broglie's waves) in which y component of electron 1 velocity in Fig.~\ref{fig:fthree} is zero at various energy levels of Helium.
WN $\times$ 4 is the total number of de Broglie's waves contained in one round of the orbital.}

\begin{ruledtabular}
\begin{tabular}{cccc}

E (eV) & r1 (MM) & WN & WN $\times$ 4 \\
\hline
 -77.00 & 3154.0 & 0.25324 & 1.01296 \\
 -77.50 & 3134.0 & 0.25242 & 1.00968 \\
 -78.00 & 3114.0 & 0.25161 & 1.00644 \\
 -78.50 & 3094.0 & 0.25080 & 1.00320 \\
 -79.00 & 3074.5 & 0.25001 & 1.00004 \\
 -79.005 & 3074.0 & 0.25000 & 1.00000 \\
 -79.01 & 3073.8 & 0.24999 & 0.99996 \\
 -79.50 & 3055.0 & 0.24922 & 0.99688 \\
 -80.00 & 3036.0 & 0.24844 & 0.99376 \\
 -80.50 & 3017.0 & 0.24767 & 0.99068 \\
 -81.00 & 2998.5 & 0.24690 & 0.98760 \\
\end{tabular}
\end{ruledtabular}
\end{table}

\begin{figure}
\includegraphics[width=6cm]{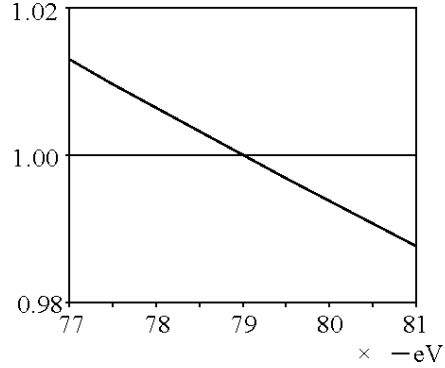}
\caption{\label{fig:ffour} Plots of the number of de Broglie's waves contained in one orbital  at various energy levels of Helium}
\end{figure}

Table~\ref{tab:table1} shows the results in which the last $VY$ is the closest to zero.
Fig.~\ref{fig:ffour} graphs the results in Table I. As shown in Table I and Fig.~\ref{fig:ffour}, when the total energy of the helium ($E$) is -79.005 eV (the experimental value),  WN $\times$ 4 is just 1.00000. 
This results demonstrate that two electrons of the helium are actually moving around the nucleus on the orbitals of just one de Broglie's wavelength as shown in Fig.~\ref{fig:ftwo} and \ref{fig:fthree}. 

Experimentally, the spin magnetic moment ($\mu$) is known to be
\begin{equation}
\mu=-\frac{g \beta S}{\hbar} \hspace{1cm} , \beta=\frac{e\hbar}{2m}
\label{eq:enine}
\end{equation}
where $g$, the spin g-factor, is 2 \cite{rtwelve}, $S$ is $\pm \frac{1}{2} \hbar$ spin angular momentum, and $\beta$ is the Bohr magneton (= $9.27400949 \times 10^{-24} \frac{J}{T}$).
If we suppose that the orbital angular momentum ($S2$) of the electron in the ground state of the hydrogen atom is $\pm \hbar$ (here, spin is zero) as in the Bohr's model and its g factor ($g2$) is 1, the magnetic moment of the electron  $\mu 2=-(g2) \beta (S2)/\hbar$ is equal to   $\mu$ of  Eq.~(\ref{eq:enine}). In this state, the above-mentioned problems of infinitely negative Coulomb potential and spinning sphere speed do not occur. As the magnetic moment of the electron is not changed, the splitting pattern of the energy levels in the magnetic field (Paschen-Back Effect) is also not changed.

In this study, the two main forces that affect the electronic motion are supposed to be the Coulomb force and the force by de Broglie's waves. By a simple calculation, the magnetic force between the two electrons is too small in comparison to the Coulomb force. In the standard model, as the helium atom has spin-up and spin-down electrons, it seems to generate no magnetic field. But to be precise, in all areas except in the part at just the same distance from the two electrons, magnetic fields are theoretically produced by the electrons even in the standard helium model. So as the electrons moves, they lose energy by emitting electromagnetic waves. 
The same thing can be said about this new helium model. In this model, the length of both orbitals crossing perpendicularly is just one de Broglie's wavelength. Probably the stability of de Broglie's waves are related to the stability of the electronic motions.

It is a very meaningful and astonishing thing that the calculation value by this new method based on the Bohr's model is just equal to the experimental value. As the Bohr's model applies to the hydrogenlike ion He$^{+}$, the second ionization energy (54.418 eV) of the helium is correctly calculated using the Borh's theory.
So, the first ionization energy (24.587 eV) can also be gotten by the equation 79.005 eV - 54.418 eV = 24.587 eV. This fact means the atomic electrons are actually moving around the nucleus, not existing as electronic clouds. it is possibile that  the basic conception of the quantum mechanics such as the uncertainty principle will be greatly changed.

\vspace{0.5cm}
Supplementary Methods; JAVA computer program to calculate the electronic orbital.

\clearpage

{\bfseries Supplementary Methods; JAVA computer program to calculate the electronic orbital.}

In this program, we first input the initial x-coordinate r1 (in MM) of electron 1, and the absolute value of the total energy E (in eV) of Helium. From the inputted values, this program outputs the y component of electron 1 velocity in Fig. 3, and WN (the number of de Broglie's waves contained in one quarter of the orbital.

\vspace{1cm}

import java.util.Scanner;

class MathMethod \{
 
public static void main (String[ ] args) \{
 
\vspace{0.5cm}
 Scanner stdIn = new Scanner ( System.in ) ; \hspace{1cm}  // input r1 and \textbar E \textbar
 
 System.out.println ("r1 between nucleus and electron 1 (MM)? ");  
 
 double r = stdIn.nextDouble ();
 
 System.out.println ("total energy \textbar E \textbar in the herium (eV) ? ");  
 
 double E = stdIn.nextDouble ();
 
 double me = 9.1093826e-31 ; double mp = 1.67262171e-27;
 
 double mn = 1.67492728e-27 ; double nucle = 2*mp + 2*mn;
   
 \hspace{7cm}                 //rm = reduced mass of an electron
 
 double rm = (2*me*nucle) / (2*(2*me+nucle));
 
 double pai = 3.141592653589793; double epsi = 8.85418781787346e-12;
 
 double h = 6.62606896e-34; double ele = 1.60217653e-19;

 for (int i = 1; i \textless 100; i++ ) \{  \hspace{2cm}    // repeat until r1=initial r1+100
                                
   \hspace{6cm}                  // calculation of initial VY from E and r1    

double poten = - (2*ele*ele*2) / (4*pai*epsi*r) + (ele*ele) / (4*pai*epsi*2*r);
                             
  \hspace{6cm}                    // vya = total E-potential energy  

double vya = - (E*1.60217646e-19) - poten*1.0e14; 
 
 if (vya \textgreater 0) \{
   
 \hspace{6cm}                     // vyb=velocity from kinetic energy

 double vyb = Math.sqrt(vya/me); 
 
 double VY = vyb*1.0e-8;   \hspace{3cm}   // change m/sec to MM/SS
 
 double prexx = r; double VX = 0.0; double WN = 0.0; double preyy=0.0; 
 
 double xx, yy, vk, preVY, preWN, midWN;
  
 do \{
    
   xx = prexx + VX; yy = preyy + VY;  \hspace{1cm}    //electron 1 position after 1SS
    
  preVY = VY; preWN = WN ;
    
  vk = VX * VX + VY * VY; \hspace{1.5cm}     // calculation of WN from VX,VY 
                        
      \hspace{7cm}             // WN = WN + Eq(8)
    
   WN = WN + (rm*vk*1.0e-6) / h;                        
   
      \hspace{4.5cm}            // calculation of VX,VY from Coulomb force
    
    double ra = Math.sqrt (prexx * prexx + preyy * preyy);       
    
    double rb = Math.sqrt (4.0 * prexx * prexx + 2.0 * preyy * preyy);
    
      \hspace{7cm}          // change MM to meter
   
    ra = ra * 1.0e-14; rb = rb * 1.0e-14; 
    
    prexx = prexx * 1.0e-14; preyy = preyy * 1.0e-14;
    
    double ac = (2 * ele * ele) / (4 * pai * epsi * rm);
    
      \hspace{7cm}       // VX = VX + Eq(5) * 10$^{-30}$
    
    VX = VX + 1.0e-30 * ac * prexx * (-1.0 / (ra*ra*ra) + 1.0 / (rb*rb*rb));   
    
      \hspace{7cm}           // VY = VY + Eq(6) * 10$^{-30}$
   
    VY = VY + 1.0e-30 * ac * preyy * (-1.0 / (ra*ra*ra) + 0.5 / (rb*rb*rb));
    
    prexx = xx; preyy = yy;
  
   \} while ( xx \textgreater 0 );    \hspace{1.5cm}    //repeat above until electron 1 arive at y axis 
   
  if (VY \textgreater -0.0001 \&\& VY \textless 0.0001) \{  \hspace{1.5cm} // last VY condition

  System.out.print ("r1: " + r + "   ");
  
  System.out.printf ("VX: \% .5f   ", VX);
  
  System.out.printf ("VY: \% .5f   ", VY);
  
  System.out.printf ("preVY: \% .5f  ", preVY);
  
  midWN = (preWN + WN) / 2; System.out.printf ("midWN: \% .5f \textbackslash n", midWN);
    
 \}
   
  \}  r = r + 1;
   
  \} \} \}


\begin{thebibliography}{00}
\bibitem{rone} N. Bohr, Philos. Mag. {\bfseries 26,} 1 (1913).

\bibitem{rtwo} C. Davisson and L.H. Germer, Nature {\bfseries 119,} 558 (1927).

\bibitem{rthree} A. Tonomura, J. Endo, T. Matsuda, T. Kawasaki and H. Ezawa, Am. J. Phys. {\bfseries 57,} 117 (1989).

\bibitem{rfour} A. A. Svidzinsky, M. O. Scully and D. R. Herschbach, Proc. Natl. Acad. Sci. U.S.A. {\bfseries 102,} 11985 (2005).

\bibitem{rfive} A. A. Svidzinsky, M. O. Scully and D. R. Herschbach, Phys. Rev. Lett. {\bfseries 95,}  080401 (2005).

\bibitem{rsix} G. Chen, Z. Ding, S-B Hsu, M. Kim and J. Zhou, J. Math. Phys. {\bfseries 47,} 022107 (2006).

\bibitem{rseven} G. E. Uhlenbeck and S. A. Goudsmit, Nature {\bfseries 117,} 264 (1926).

\bibitem{reight} C. L. Pekeris, Phys. Rev. {\bfseries 115,} 1216 (1959).
 
\bibitem{rnine} C. W. Scherr and R. E. Knight, Rev. Mod. Phys. {\bfseries 35,} 436 (1963).

\bibitem{rten} R. Ahlrichs, Phys. Rev. A {\bfseries 5} 605 (1972). 

\bibitem{releven} J. D. Baker, D. E. Freund, R. N. Hill and J. D. Morgan, Phys. Rev. A {\bfseries 41,} 1247 (1990).

\bibitem{rtwelve} G. Gabrielse, D. Hanneke, T. Kinoshita, M. Nio and B. Odom, Phys. Rev. Lett. {\bfseries 97,} 030802 (2006).

\end{thebibliography}
\end{document}